\documentclass[onecolumn,floatfix,ams,nofootinbib]{revtex4}
\usepackage[dvips]{epsfig}
\usepackage[english]{babel}
\usepackage{bbm}
\usepackage{verbatim}
\usepackage{array}
\usepackage{float}
\usepackage{bm} 
\usepackage{amsmath}
\usepackage{amsfonts}
\usepackage{amssymb}
\usepackage{hyperref}
\usepackage[dvipsnames]{xcolor}
\hypersetup{colorlinks=true,linkbordercolor=Red,linkcolor=Red, citecolor=Red}
\usepackage{indentfirst} 
\usepackage{float}
\usepackage{graphicx} 
\usepackage{lipsum}		    
\usepackage{amssymb, amsmath, amsthm}
\usepackage{mathrsfs}
\usepackage{dsfont}

\begin{document}

\title{Non-standard Wigner doublets}

	\author{F. A. da Silva Barbosa} 
\email{felipe.barbosa@edu.ufes.br}
\affiliation{N\'ucleo Cosmo-ufes - Departamento de Qu\'imica e F\'isica,
	Universidade Federal do Esp\'irito Santo - Campus Alegre, ES, 29500-000, Brazil.}

\author{J. M. Hoff da Silva} 
\email{julio.hoff@unesp.br}
\affiliation{Departamento de F\'isica, Universidade
	Estadual Paulista, UNESP, Av. Dr. Ariberto Pereira da Cunha, 333, Guaratinguet\'a, SP,
	Brazil.}

\begin{abstract}
Guided by a conservative formulation in investigating the physical content of quantum fields, we explore non-standard Wigner classes of particles that could provide the basis for self-interaction models to dark matter. We critically contrast the analysis with long-standing constraints to non-standard Wigner classes in the literature to discuss the model's viability.   
\end{abstract}

\maketitle

\section{Introduction}

An epoch-made paper \cite{EPW} showed that the quantum labels characterizing a particle come from space-time symmetries represented in the Hilbert space. These labels are the well-known concepts of mass and spin (or helicity for zero-mass particles). From the formulation of how these particles interact and how to adequately describe it in a covariant fashion emerges the very concept of a quantum field and its theory \cite{weinberg_1995}, which, in turn, gives the theoretical scope for the so-called standard model conceptualization.   

The search for dark matter is one of the most interesting open problems in current high-energy physics. Several attempts to describe dark matter dynamics and particle content have appeared from the theoretical counterpart in the literature. In a broad brush, these attempts run from exotic models, those violating one or more symmetries, to more conservative approaches. By turn, it is convenient to separate the conservative candidates into two classes: models composed by fields whose particle content is entirely determined by the orthochronous proper Lorentz subgroup labels, including higher spin theories to standard fields interacting weekly, and models exploring further particle degeneracy coming from the inclusion of space-time reflections into the particle representation scheme. The present paper deals with possibilities coming from this last scenario.

It is curious that, although the robustness of the approach presented in \cite{EPW}, discrete symmetries were not correctly investigated\footnote{The time reversal operator was taken as acting in a unitary way, rendering the non-physical possibility of negative energies irreducible representations.}. This problem was corrected \cite{Wigner,weinberg_1995}, and it was envisaged that additional degeneracy particle labels could be in order in a more comprehensive theory. Fermionic possibilities coming from this more general scope were extensively treated in \cite{ag}, and more recently, a doubling of degrees of freedom along with Wigner degeneracy were studied in the context of dark matter candidates \cite{ach2} (see also \cite{ach} for technical details in a previous version of the theory). In this paper, we critically study states endowed with additional degeneracy, the Wigner doublets, and their corresponding fields in the general $SU(2)\otimes SU(2)$ form ($(A,B)$ fields), investigating how interacting models may be constructed, circumventing a general criticism regarding non-standard Wigner classes \cite{Lee_Wick}, and bearing in mind a connection with a theoretical dark matter formulation. As we are going to see, severe assumptions must be elicited and their physical constraints respected in order to get a viable model.   

This work is organized as follows: after a quick review of the basic formalism, we depict the states composing the Wigner doublets in the context addressed here in section II. These states are to be understood as the quantum of general $(A,B)$ fields introduced in section III, also devoted to presenting the behavior of the fields under reflections. Section IV discusses interaction models, framing the discussion in the context of theoretical dark matter candidates. When taking all the necessary constraints, the models' viability points to a narrow possibility, as we shall see. In the final section we conclude.   

\section{The basic setup to Wigner doublets}

We shall depict the basic setup following Ref. \cite{weinberg_1995} to standardize the notation. Our starting point is the exponential series representation for the S-matrix \cite{PhysRev.75.1736},
 \begin{align}\label{Dysonform}
     \mathbf{S}  =  \mathbf{1}  + \sum_{n=1}^{\infty} \frac{(-i)^n}{n!} \int d^4 x_1 \cdots d^4 x_n T\{\mathcal{H}(x_1) \cdots \mathcal{H}(x_n)\},
 \end{align} 
where $\mathcal{H}(x)$ is the interaction Hamiltonian density in the interaction picture\footnote{Given $H = H_0 + V$, then $V(t) = e^{i H_0 t}\, V \,e^{- i H_0  t} \equiv \int d^3 x \, \mathcal{H}(x)$.} and $T$ is the time ordering operator. The $S$ matrix will be perturbatively Lorentz invariant if 
\begin{subequations}\label{SmatrixLI}
\begin{align}
   &U_0 ( \Lambda, a) \mathcal{H}(x) U^{-1}_0 ( \Lambda,a) = \mathcal{H}(\Lambda x + a) \quad \text{and} \label{eq:suba} \\
   &\left[ \mathcal{H}(x)  ,  \mathcal{H}(y)  \right] = 0, \quad \text{for $(x-y)^2 \geq 0  $}, \label{eq:subb}
\end{align}
\end{subequations}
where $U_0 ( \Lambda,a)$ is a unitary operator performing Lorentz transformations on free particle states. In addition to being Lorentz invariant, one also expects the S-matrix to satisfy the cluster decomposition principle \cite{PhysRev.132.2788} stating that distant experiments must produce uncorrelated physical results. It can be shown \cite{weinberg_1995} that this will be the case if\footnote{Here $q = (\boldsymbol{p},\sigma,n)$, with $(\sigma,n)$ denoting spin and particle type and $a^{\dagger}(q)$ is the free particle creation operator.}
\begin{align}\label{CDP}
    V = \sum_{N=0}^{\infty} \sum_{M=0}^{\infty}  \int \left( \prod_{i=1}^{N} d q'_i \right) \left( \prod_{j=1}^{M} d q_j \right) a^{\dagger} (q'_1) \dots a^{\dagger} ( q'_N)  \, a(q_1) \dots a (q_M) \times \nonumber \\
    \times \delta^3 (\boldsymbol{p}'_1 + \dots + \boldsymbol{p}'_N - \boldsymbol{p}_1 - \dots -\boldsymbol{p}_M ) \, V_{NM} ( q'_1 \dots q'_N, q_1 \dots q_M ),
\end{align}
with $V_{NM} ( q'_1 \dots q'_N, q_1 \dots q_M )$ being smooth in the sense that it does not contain any delta functions of momenta.

In order to satisfy \eqref{eq:suba} $\mathcal{H}(x)$ is built out of fields $\psi_n (x)$ so that
\begin{align}\label{Field definition}
   U_0 ( \Lambda, a) \psi_n(x) U^{-1}_0 ( \Lambda,a) =  \sum_m D_{n m } (\Lambda^{-1} ) \psi_m (\Lambda x + a),
\end{align}
where $D_{n m }$ is a matrix representation of the Lorentz group; the validity of \eqref{CDP} is guaranteed if one uses creation and annihilation fields\footnote{The dependence of the fields on $e^{i p \cdot x}$ is set by \eqref{Field definition} with $(\Lambda,a) = (\mathbf{1},a)$ and the form of $u_n ( \boldsymbol{p} , \sigma)$ and $v_n ( \boldsymbol{p} , \sigma)$ is fixed by the choice of matrix representation $D_{mn}$. }:
\begin{align}
   &\psi^{+}_n (x) = (2 \pi )^{-3/2} \sum_\sigma \int d^3 \boldsymbol{p} \, e^{i p \cdot x } \, u_n ( \boldsymbol{p} , \sigma) \, a( \boldsymbol{p}, \sigma ), \\
    &\psi^{-}_n (x) = (2 \pi )^{-3/2} \sum_\sigma \int d^3 \boldsymbol{p} \, e^{-i p \cdot x } \, v_n ( \boldsymbol{p} , \sigma) \, a^{\dagger} ( \boldsymbol{p}, \sigma ).
\end{align}
By grouping these fields into Lorentz invariant products in the correct order, $\psi^{-}_n (x)$ to the left of $\psi^{+}_n (x)$, \eqref{eq:suba} and \eqref{CDP} are automatically satisfied. However, an arbitrary Lorentz invariant product of $\psi^{-}_n (x)$ and $\psi^{+}_n (x)$ will not satisfy \eqref{eq:subb}, since their commutator (or anticommutator) is in general different than zero for $(x-y)^2 \geq 0 $. The solution is to work with linear combinations of creation and annihilation fields which commute (or anticommute) at space-like separations:
\begin{subequations}\label{localfields}
\begin{align}
    &\psi_n (x) = k \psi^+_n (x) + \lambda \psi^{-}_n (x),\label{local:a} \\
    &\left[ \psi_n (x), \psi_m (y) \right]_{\mp} = \left[ \psi_n (x), \psi^{\dagger}_m (y) \right]_{\mp} = 0. \label{local:b}
\end{align}
\end{subequations}
Since $\mathcal{H}(x)$ is a product of these fields\footnote{Notice that it has to be an even polynomial of fields which anticommute.} it will satisfy \eqref{eq:subb}. It is in the construction of fields that satisfy \eqref{localfields} that one can prove the spin-statistics theorem for $\mathcal{H}^{\dagger}(x) = \mathcal{H}(x)$. The introduction of \eqref{localfields} is enforced by \eqref{eq:subb}, but a viable $\mathcal{H}(x)$ can not be a simple product of  $\psi_n (x)$ since these would violate \eqref{CDP}. As it is well known, to simultaneously ensure Lorentz invariance and cluster decomposition, it is necessary to construct Lorentz invariant combinations in the so-called normal ordering: any product of fields is to be understood with all the creation operators to the left of annihilation operators, ignoring non-vanishing commutators. Moreover, if we have particles that carry conserved quantum numbers, $Q \, | \boldsymbol{p},\sigma, n \rangle = q \, |\boldsymbol{p},\sigma,n \rangle $ and $\left[Q,H\right]=0$, one has to define $\psi_n(x)$ as a sum of particle and anti-particle fields\footnote{The anti-particle being defined as a particle with the same mass and spin but opposite quantum number: $Q \, | \boldsymbol{p},\sigma, n^c \rangle = -  q \, |\boldsymbol{p},\sigma,n^c \rangle $.}
\begin{align}\label{complexfield}
    \psi_m (x) = \sum_\sigma\int \frac{d^3 \boldsymbol{p}}{ (2 \pi )^{3/2}} \left[k \, e^{i p \cdot x } \, u_m ( \boldsymbol{p} , \sigma, n) \, a( \boldsymbol{p}, \sigma ,n ) + \lambda\,   e^{- i p \cdot x } \, v_m ( \boldsymbol{p} , \sigma, n^c) \, a^{\dagger}( \boldsymbol{p}, \sigma ,n^c)  \,  \right],
\end{align}
which will have the simple commutator $\left[ Q ,\psi_m (x)\right] = - q \psi_m (x)$, so that $\mathcal{H}(x)$ will commute with $Q$ if it is constructed out of products $\psi(x) \, \psi^{\dagger} (x)$.

In addition to providing a rationale for quantum field theory, the formalism sketched above allows the derivation of practical results in quantum field theory, like the relations of phases under space-time inversions, for any spin in a concise way. Therefore, it seems the appropriate framework to investigate new possibilities. Naturally, the conclusions obtained in this way are limited by the constraints we impose, that is  \eqref{SmatrixLI}, \eqref{CDP} and $\mathcal{H}^{\dagger}(x) = \mathcal{H}(x)$. We will deal in the rest of this work with $(A,B)$ fields. These are irreducible representations of the proper Lorentz group, that is
\begin{align}\label{abfields}
U_0 ( \Lambda) \psi^{AB}_{ab}(x) U^{-1}_0 ( \Lambda) =  \sum_{a' \, b'} D^{A0}_{a,a' } (\Lambda^{-1} ) D^{0B}_{b,b' } (\Lambda^{-1} ) \psi^{AB}_{a' \, b'} (\Lambda x ),
\end{align}
where $D^{A0}_{a,a' } (\Lambda^{-1} ) \text{and } D^{0B}_{b,b' }(\Lambda^{-1} )$ are, respectively, the irreducible matrix representations $(A,0)$ and $(0,B)$ of the proper group. $A$ and $B$ can only be integers or half integers, and an $(A,B)$ field will be constructed with the creation and annihilation operators of a spin $j$ particle, where $j= A+B, A+B -1, \dots ,|A-B|$.

\subsection{Wigner doublets}\label{Wigner section}

As it is well known, quantum field theory describes the interactions of particles which, under space-time inversions, do not change particle type. Under time reversal, the state of an electron transforms into a new state with different spin components and momenta while still representing an electron. Nonetheless, the existence of more general possibilities has been recognized long ago \cite{Wigner}, in the context of degeneracy beyond the spin. In other words, time reversal can exhibit non-trivial behavior within a particle multiplet\footnote{The label $n$ now denotes a set of particles with the same mass and spin.}
\begin{align}\label{generalact}
    T \, |\boldsymbol{p} , \sigma , n  \rangle = (-1)^{j-\sigma}  \sum_{m} \mathcal{T}_{mn} |-\boldsymbol{p},-\sigma,m \rangle,
\end{align}
with $\mathcal{T}_{mn}$ an arbitrary unitary matrix. Since time reversal is an antiunitary transformation, it is not possible to diagonalize $\mathcal{T}_{mn}$ by simply redefining the basis states. However, it is possible to transform it into a block-diagonal form, where the blocks can be either $1\times 1$ unity matrices or $2\times 2$ matrices with the following form \cite{weinberg_1995}
\begin{center}
$ \begin{pmatrix}
0 & e^{i \phi_l/2} \\
e^{-i \phi_l/2} & 0
\end{pmatrix}$,
\end{center}
where $e^{i \phi_l} \neq 1$. Hence, the general action of time reversal on a multiplet includes particles that transform conventionally and doublets that transform among themselves.

This degeneracy under $T$ can be physical only in cases where no internal symmetry operators exchange particles from the doublet into each other. In particular, consider just one doublet $| \boldsymbol{p},\sigma,\pm \rangle$, so that $T \, | \boldsymbol{p},\sigma,\pm \rangle = (-1)^{j-\sigma} e^{\pm i \phi/2} | - \boldsymbol{p},- \sigma,\mp \rangle $. If there is a unitary symmetry operator $S$ whose action is
\begin{align}\label{S symmetry}
    S \, | \boldsymbol{p},\sigma,\pm \rangle =  e^{\pm i \phi/2} | \boldsymbol{p},\sigma,\mp \rangle, \quad \text{and} \quad
    S \, | \boldsymbol{p},\sigma, n  \rangle = \eta^n | \boldsymbol{p},\sigma, n  \rangle,
\end{align}
where $| \boldsymbol{p},\sigma, n  \rangle$ denotes particles with diagonal transformation properties under $T$, one can define a new operator $T' \equiv S^{-1} T $, so that under time reversal, all particles transform in the usual way. In other words, this internal symmetry operator can not exist in a non-trivial theory of time-reversal doublets. This point is a relevant consistency constraint and will be further explored in the sequel.

On the same multiplet where time reversal acts nontrivially, one can also have an unusual action of parity
\begin{align}\label{PARITY}
    P\, |\boldsymbol{p} ,\sigma ,n  \rangle = \sum_m \mathcal{P}_{mn} |-\boldsymbol{p},\sigma, m \rangle,
\end{align}
where $\mathcal{P}_{mn}$ is a unitary matrix. Since parity is a unitary transformation, $\mathcal{P}_{mn}$ can always be made diagonal. However, the choice of basis where it is diagonal is not necessarily the same where $\mathcal{T}_{mn}$ is block diagonal. We will deal with just one doublet here, and parity considerations will be made once we construct $\mathcal{H}(x)$ in the next section. We see three possibilities for the particle-antiparticle degeneracy of time reversal doublets: the antiparticle of $| \boldsymbol{p},\sigma,\pm \rangle$ could either be part of the doublet, $| \boldsymbol{p},\sigma,\mp \rangle$, or not. We shall investigate the first case in the following sections.

\section{General $(A,B)$ fields}

Since we are keeping the possibility that the anti-particle of $| \boldsymbol{p},\sigma,\pm \rangle$ is $| \boldsymbol{p},\sigma,\mp \rangle$ we take the $(A,B)$ fields \eqref{ABfield} to have the form \eqref{complexfield}
\begin{align}
\psi^{AB}_{ab}  = ( 2 \pi)^{-3/2} \sum_{\sigma} \int d^3 p \left[ k \, e^{i p \cdot x }\, u^{AB}_{ab}(\boldsymbol{p}, \sigma) \, a(\boldsymbol{p}, \sigma) + \lambda \, e^{-ip \cdot x} v^{AB}_{ab}(\boldsymbol{p}, \sigma)\, a^{c \dagger}(\boldsymbol{p},\sigma) \right],
\end{align} for a spin $j$ doublet, with $j = A+B , A+B - 1, \dots , |A-B|$, and where $a^{\dagger}(\boldsymbol{p}, \sigma ) \equiv a^{\dagger}( \boldsymbol{p}, \sigma, +)$ and $a^{c \dagger} (\boldsymbol{p}, \sigma) \equiv a^{\dagger}(\boldsymbol{p}, \sigma, -)$ for comparison with \cite{weinberg_1995}. The restriction of \eqref{abfields} to pure spatial rotations shows that $v^{AB}_{ab}(\boldsymbol{0},\sigma) = (-1)^{j+\sigma} u^{AB}_{ab}(\boldsymbol{0}, - \sigma) $ and $u^{AB}_{ab}(\boldsymbol{0}, \sigma)$ is, up to a proportionality constant, the Clebsch-Gordon coefficient $C_{AB}(j \sigma ; ab)$ which combines $(A,B)$ spin states to form a state of spin $j$. By adopting the proportionality constant to be $(2 m )^{-1/2}$ we obtain
\begin{align}
    &u^{AB}_{ab}(\boldsymbol{0}, \sigma) = (2m)^{-1/2} C_{AB}(j \sigma; ab),\\
    &u^{AB}_{ab}(\boldsymbol{p}, \sigma) = \frac{1}{\sqrt{2 p^0}} \sum_{a' b'} \left( \exp{\left(- \hat{\boldsymbol{p}} \cdot  \boldsymbol{J}^{(A)} \theta \right) } \right)_{a a'} \left( \exp{ \left( \hat{\boldsymbol{p}} \cdot \boldsymbol{J}^{(B)} \theta \right) } \right)_{b b'} C_{AB}(j \sigma; ab),\\
    &v^{AB}_{ab}(\boldsymbol{p}, \sigma) =  (-1)^{j+\sigma}  u^{AB}_{ab}(\boldsymbol{p}, - \sigma),
\end{align}
where $\boldsymbol{J}^{(A)}$ and $\boldsymbol{J}^{(B)}$ are spin $A$ and $B$ representations of angular momenta and $\theta$ is defined by $\text{cosh}(\theta) = \sqrt{\boldsymbol{p}^2  + m^2}/m$ and $\text{sinh}(\theta)  = | \boldsymbol{p}|/m$.

Since our analysis is fairly connected to the usual formulation, conclusions relying only on invariance considerations under the orthochronous group are immediately inherited. In particular, to satisfy \eqref{local:b}, that is $ \left[ \psi^{AB}_{ab} (x) , \psi^{\dagger AB}_{\tilde{a} \tilde{b}} (y) \right]_{\mp}  = 0$ for $(x-y)^2 \geq 0$, it is necessary that $|k| =| \lambda |$ and $\pm (-1)^{2j}  =1 $, where $\pm$ stands for $\mp$ on $[\cdot,\cdot]_{\mp}$. In other words, integer spins are bosons and half-integer spins are fermions. Therefore, after properly rescaling the field, we have 
\begin{align}
\psi^{AB}_{ab}  = ( 2 \pi)^{-3/2} \sum_{\sigma} \int d^3 p \left[  e^{i p \cdot x }\, u^{AB}_{ab}(\boldsymbol{p}, \sigma) \, a(\boldsymbol{p}, \sigma) + (-)^{2B} c \, e^{-ip \cdot x} v^{AB}_{ab}(\boldsymbol{p}, \sigma)\, a^{c \dagger}(\boldsymbol{p},\sigma) \right],
\end{align}
where $c$ is a phase factor independent of the $(A,B)$ representation. It is straightforward to show that 
\begin{align}
T \psi^{AB}_{ab}(x) T^{-1} = (-)^{A+3B+j} \frac{ e^{-\frac{i \phi}{2} } }{(2 \pi)^{3/2}} \sum_{\sigma} \int d^3 p \Big[  c^* \, e^{-i p \cdot Tx} \, u^{* BA}_{ba} (\boldsymbol{p}, \sigma) \, a^{\dagger}(\boldsymbol{p}, \sigma) \nonumber\\
+ (-)^{2A} \, e^{i p \cdot Tx } \, v^{* BA}_{ba}(\boldsymbol{p}, \sigma)\, a^{c} (\boldsymbol{p}, \sigma) \Big].
\end{align}
To ensure that $ T \psi^{AB}_{ab}(x) T^{-1} \propto \psi^{\dagger BA}_{ba} (Tx)$, it is necessary that $c= \pm 1$. If $c=-1$, we can redefine the one-particle states so that\footnote{Since $T$ acts on Wigner doublets in the same way that $CT$ acts on usual particle-antiparticle doublets, the same transformation can be obtained from the action of $CT$ on $(A,B)$ fields of particles which have the usual transformation under $T$.} 
\begin{align}\label{ABfield}
\psi^{AB}_{ab}  = ( 2 \pi)^{-3/2} \sum_{\sigma} \int d^3 p \left[  e^{i p \cdot x }\, u^{AB}_{ab}(\boldsymbol{p}, \sigma) \, a(\boldsymbol{p}, \sigma) + (-)^{2B} \, e^{-ip \cdot x} v^{AB}_{ab}(\boldsymbol{p}, \sigma)\, a^{c \dagger}(\boldsymbol{p},\sigma) \right],
\end{align}
and
\begin{align}\label{Ttransformation}
T \psi^{AB}_{ab}(x) T^{-1} = e^{-i \frac{\tilde{\phi}}{2}} (-)^{A+3B+j} \psi^{\dagger BA}_{ba} (Tx),
\end{align}
where $ e^{\frac{-i \tilde{\phi}}{2}} \equiv - e^{\frac{-i \phi}{2}}$. From now on, we make $\tilde{\phi} \rightarrow \phi$.

In the section \ref{Wigner section}, the behavior of doublets under parity \eqref{PARITY} was left arbitrary. We now attempt to restrict the matrix $\mathcal{P}_{mn}$ by demanding that space inversion also acts in a covariant way. At first glance, we have  
\begin{align*}
&P \psi^{AB}_{ab}(x) P^{-1} = \\
&(2 \pi)^{-3/2} \sum_\sigma \int d^3 p \left[ e^{i p \cdot Px }\, u^{AB}_{ab}(- \boldsymbol{p}, \sigma)\, \mathcal{P}^*_{++} \, a(\boldsymbol{p}, \sigma) + (-)^{2B} \, e^{-ip \cdot Px} v^{AB}_{ab}( -\boldsymbol{p}, \sigma)\,\mathcal{P}_{--} \, a^{c \dagger}(\boldsymbol{p},\sigma) \right] \\
+ &(2 \pi)^{-3/2} \sum_\sigma \int d^3 p \left[ e^{i p \cdot Px }\, u^{AB}_{ab}(- \boldsymbol{p}, \sigma)\, \mathcal{P}^*_{+-} \, a^c(\boldsymbol{p}, \sigma) + (-)^{2B} \, e^{-ip \cdot Px} v^{AB}_{ab}( -\boldsymbol{p}, \sigma)\,\mathcal{P}_{-+} \, a^{ \dagger}(\boldsymbol{p},\sigma) \right].  
\end{align*}
By using 
\begin{align*}
& u^{AB}_{ab}(- \boldsymbol{p}, \sigma) = (-)^{A+B-j} \,  u^{BA}_{ba}(\boldsymbol{p}, \sigma), \qquad \qquad v^{AB}_{ab}(-\boldsymbol{p}, \sigma) = (-)^{A+B-j} \, v^{BA}_{ba}( \boldsymbol{p}, \sigma), \\
& u^{AB}_{ab}(- \boldsymbol{p}, \sigma) = (-)^{A+B+a+b} \, v^{* AB}_{-a -b}(\boldsymbol{p}, \sigma), \qquad v^{AB}_{ab}(- \boldsymbol{p},\sigma) = (-)^{A+B+a+b+2j} \, u^{*AB}_{-a-b}(\boldsymbol{p}, \sigma),
\end{align*}
it follows that
\begin{align*}
&P \psi^{AB}_{ab}(x) P^{-1} = \\
&\frac{ (-)^{A+B-j}}{(2 \pi)^{3/2}} \sum_\sigma \int d^3 p \left[ e^{i p \cdot Px }\, u^{BA}_{ba}( \boldsymbol{p}, \sigma)\, \mathcal{P}^*_{++} \, a(\boldsymbol{p}, \sigma) + (-)^{2B} \, e^{-ip \cdot Px} v^{BA}_{ba}( \boldsymbol{p}, \sigma)\,\mathcal{P}_{--} \, a^{c \dagger}(\boldsymbol{p},\sigma) \right] +\\
&\frac{(-)^{A+3B+a+b}}{(2 \pi)^{3/2}} \sum_\sigma \int d^3 p \left[ (-)^{2B} \,e^{i p \cdot Px }\, v^{* AB}_{-a-b}( \boldsymbol{p}, \sigma)\, \mathcal{P}^*_{+-} \, a^c(\boldsymbol{p}, \sigma) + (-)^{2j} \, e^{-ip \cdot Px} u^{*AB}_{-a-b}( \boldsymbol{p}, \sigma)\,\mathcal{P}_{-+} \, a^{ \dagger}(\boldsymbol{p},\sigma) \right].
\end{align*}
For $P$ to have a covariant action, it is necessary that
\begin{align}
\mathcal{P}_{--} = \mathcal{P}^*_{++} (-)^{2j}, \qquad \mathcal{P}_{-+} = \mathcal{P}^*_{+-} (-)^{2j},
\end{align}
in which case
\begin{align}
P \psi^{AB}_{ab}(x) P^{-1} = (-)^{A+B-j} \, \mathcal{P}^{*}_{++} \,  \psi^{BA}_{ba}(Px)\,  +\,  (-)^{A+3B+a+b} \,  \mathcal{P}^{*}_{+-} \,  \psi^{\dagger AB}_{-a-b}(Px).
\end{align}
On the other hand, considering the unitarity of $\mathcal{P}_{mn}$, i.e., $|\mathcal{P}_{++} |^2 + | \mathcal{P}_{+-}|^2 = 1$ and $\mathcal{P}_{++} \, \mathcal{P}_{+-} = 0$, one is forced to conclude that $\mathcal{P}_{mn}$ can only be diagonal or anti-diagonal to this case. The fields can transform in either of the following ways\footnote{As in the time reversal case, \eqref{parityB} can be derived from the action of $CP$ on the $(A,B)$ fields of usual particles.}:
\begin{subequations}
	\begin{align}
	&P \psi^{AB}_{ab}(x) P^{-1} = (-)^{A+B-j} \, \mathcal{P}^{*}_{++} \,  \psi^{BA}_{ba}(Px) , \label{parityA}\\
	&P \psi^{AB}_{ab}(x) P^{-1} =  (-)^{A+3B+a+b} \,  \mathcal{P}^{*}_{+-} \,  \psi^{\dagger AB}_{-a-b}(Px) .\label{parityB}
	\end{align}
\end{subequations}

Even though the $(A,B)$ field and its transformation properties are all we need to construct the interaction density $\mathcal{H}(x)$, as mentioned in \ref{Wigner section}, only in theories which are not invariant by the $S$ symmetry \eqref{S symmetry} that the doublet can be attributed to space-time inversions. We take the absence of this symmetry as a guiding principle in the models' discussion. Hence, to study interactions that are not invariant by $S$, it is useful to know how the operator acts at the field level. From \eqref{S symmetry} and \eqref{ABfield} we see that $S$ acting on $| \boldsymbol{p}, \sigma, \pm \rangle $  behaves like $C$ for usual particles, so that 
\begin{align}\label{psitrans}
S \psi^{AB}_{ab} (x) S^{-1} =  e^{-i \phi/2} (-)^{-2A -a -b -j} \psi^{\dagger BA}_{-b -a}(x).
\end{align}
On the other hand, since $S | \boldsymbol{p},\sigma,n \rangle = \eta  | \boldsymbol{p},\sigma,n \rangle$ for usual particles, if $\Psi^{AB}_{ab}(x)$ represents its field, then  
\begin{align}\label{psitransfusual}
S \Psi^{AB}_{ab}(x)S^{-1} = \eta^* \Psi^{AB}_{ab}(x),
\end{align}
where $\eta^* = \eta^c$. We shall return to these transformations in the next section, attempting to construct non-trivial models of inversion doublets.

\section{A discussion on Interacting models}

So far, field theory tools have been used to derive general properties of time reversal doublets. Now, we address the problem of studying non-trivial viable interacting models within the context here aborded. We shall critically discuss the modeling, imposing the physical constraints and, then, narrowing more and more the possibilities. The main point of our examples is to set a basis to discuss the properties that these non-standard interacting models are expected to satisfy and their status as dark matter candidates. Consider a scalar doublet ($A=B=0$):
\begin{align}
T  \phi(x) T^{-1} = e^{-i\phi/2} \phi^{\dagger} (Tx),
\end{align}
with the usual behavior under parity \eqref{parityA}. We will require $U(1)$ symmetry, in which case $\mathcal{H}(x)$, or equivalent the Lagrangian, depends only on powers of $\phi \, \phi^\dagger$.

Due to the $U(1)$ invariance, one may expect that $\phi$ couples to the photon at low energies. However, that is not true if we require invariance under space-time reflections. Indeed, the conserved current 
\begin{align}
j^\mu_\phi (x) \equiv i \left( \phi^\dagger \partial^\mu \phi -  \partial^\mu \phi^\dagger \phi \right)
\end{align}
has the transformation law\footnote{Ignoring a c-number commutator.} $T j^{\mu}_\phi (x) T^{-1} = - \mathcal{P}^\mu\,_\nu j^\nu_\phi (Tx)$, where $P^\mu\,_\nu$ is the parity matrix. On the other hand, the QED current transforms like $T j^{\mu}_{qed} (x) T^{-1} = + \mathcal{P}^\mu\,_\nu j^\nu_{qed} (Tx)$. There are two potentials, $A^\mu$ and $B^{\mu}$, through which matter and photons can couple \cite{QED_GR}. Under space-time reflections, these potentials have the following transformation law
\begin{align}
& P A^\mu (x) P^{-1}   =  \mathcal{P}^\mu\,_\nu A^{\nu}(Px), \qquad  \quad  P B^\mu (x) P^{-1} = - \mathcal{P}^\mu\,_\nu B^{\nu}(Px),\\
& T A^\mu (x) T^{-1}   =  \mathcal{P}^\mu\,_\nu A^{\nu}(-Px), \qquad    T B^\mu (x) T^{-1} = - \mathcal{P}^\mu\,_\nu B^{\nu}(-Px).
\end{align}
Naturally, the coupling between matter and photons is $\mathcal{H}(x) = A_\mu (x) j^{\mu}_{qed} (x)$ and therefore there is no coupling between photons and $j^{\mu}_\phi (x)$ preserving space-time reflections, since $ P j^{\mu}_\phi (x) P^{-1} = \mathcal{P}^\mu\,_\nu j^\nu_{\phi} (Px) $. Had we choose a scalar field with unusual transformation under $P$ \eqref{parityB}, then it would couple\footnote{It has been recently shown that such couplings, although not Lorentz invariant at any finite order in perturbation theory, are Lorentz invariant at the non-perturbative level \cite{LI}.} to $B^\mu$. This possibility is excluded since we are ultimately interested in dark matter candidates. 

As mentioned before, a non-trivial model of doublets can only be realized if there is no internal symmetry operator $S$ such that the time-reversal operator can be redefined to have the usual action. If we restrict ourselves to the scalar field and $U(1)$ invariance, the simplest Lagrangian which meets this requirement is 
\begin{align}\label{simplephi}
\mathcal{L}= - \frac{1}{2} \partial^\mu \phi \partial_\mu \phi^* - \frac{1}{2} m^2 \phi \phi^*  - \frac{\lambda}{M^2} \text{Re} \left[ \phi^* \partial_\mu \phi \right] i \, \text{Im} \left[ \phi^* \partial^\mu \phi \right],
\end{align}
where $M$ is some cutoff mass. The transformation \eqref{psitrans} is equivalent to $ \phi \rightarrow \phi^* $, under which the last term is not invariant. Clearly, \eqref{simplephi} is not renormalizable, not even as an effective field theory. To regularize the divergence of a $2 \rightarrow 2$ scattering, we would need a counterterm of the type $\left( \phi \phi^* \right)^2$. Then, with the addition of higher-order terms, the theory could be renormalizable as an effective field theory. However, at very low energies, where only the renormalizable terms are relevant, the model would appear trivial since the renormalizable contributions are invariant under the $S$ operator. It is a reasonable requirement for a non-trivial model of time-reversal doublets to be non-trivial at all energy scales. In this case, it is not possible to construct such a model with the scalar field alone.

We foresee two simple ways to extend the field content to construct a non-trivial, consistent model. The first relies on the addition of a massive real vector field, with the transformation properties
\begin{align}
P A^\mu P^{-1} = \eta_P \mathcal{P}^\mu\,_\nu A^\nu (Px), \qquad T A^\mu T^{-1} = \eta_T \mathcal{P}^\mu_\nu A^\nu (-Px),
\end{align}
where $\eta_P$ and $\eta_T$ are real phases. As long as $\eta_P =- \eta_T = 1$, we have the following possibility:
\begin{align}\label{poss1}
\mathcal{L} = \mathcal{L}_0 +   \lambda  A^\mu  \text{Re} \left[ \phi^* \partial_\mu \phi \right]  +  \bar{\lambda} A^\mu i \, \text{Im} \left[ \phi^* \partial_\mu \phi \right],
\end{align} where $\mathcal{L}_0$ contains the free theory. It is easy to see that this model is not invariant by $S$ since the different currents formed by the scalar field have different transformations, and the vector field transformation is \eqref{psitransfusual} $S A^\mu (x) S^{-1} = \eta A^{\mu}(x)$, with $\eta = \pm 1$. At this point, one may notice that \eqref{poss1} seems identical to an ordinary $CT-\text{invariant}$ theory, where $\phi$ has the usual transformation under $T$ and $C$, so before going to the next case we pause to consider the physical difference between both cases. Although we shall not advocate in favor the model at hand, this analysis is elucidating.

Ordinary $T$ invariance, where $T$ does not change particle content, in the language of the $S$ matrix, translates to\footnote{We omit the phase factors of the transformation for clarity.}
\begin{align}\label{T-INVA-S-MATRIX}
S_{\boldsymbol{p}_1  \sigma_1 n_1 \, ,\, \boldsymbol{p}_2 \sigma_2  n_2,\dots \, ;\,  \boldsymbol{p}'_1 \sigma'_1 n'_1 \, ,\, \boldsymbol{p}'_2 , \sigma'_2 n'_2 \dots  } &= (- )^{j_1 - \sigma_1} (-)^{j_2 - \sigma_2} \dots (-)^{j'_1 - \sigma'_1} (-)^{j'_2-\sigma'_2} \dots \times\nonumber\\
& \times S_{-\boldsymbol{p}'_1 -\sigma'_1 \, n'_1 \, ,\, -\boldsymbol{p}'_2 , -\sigma'_2 \, n'_2 \dots \, ;\, -\boldsymbol{p}_1 - \sigma_1 \,  n_1 \, ,\, - \boldsymbol{p}_2 -\sigma_2 \, n_2,\dots   } . 
\end{align}
Put simply, the probability for some initial state $\alpha$ to collide and produce $\beta$ is equal to the probability of the initial state $\beta$ to produce $\alpha$ when we invert the momenta and spin projection of each particle involved in the collision, but not the particle type. In a usual $CT$ invariant theory \eqref{T-INVA-S-MATRIX} is no longer valid. However, there is a closely related equality where, in addition to the previous transformations, we take $n_i \rightarrow n^c_i$, or $\pm q \rightarrow \mp q$ in our case. This is the same invariance of the $S$ matrix predicted by the non-trivial time reversal doublet theory \eqref{poss1}. At first sight, one may think that both theories are physically indistinguishable. Nevertheless, further investigation shows that this is not the case. In principle a $CT$ invariant theory can be reformulated to satisfy \eqref{T-INVA-S-MATRIX} by considering
\begin{align}\label{redefinition}
| \boldsymbol{p}, \sigma, \pm \rangle = \frac{1}{\sqrt{2}}\left( | \boldsymbol{p}, \sigma, + q \rangle  \pm  | \boldsymbol{p}, \sigma, - q \rangle \right),
\end{align}
so that $CT  | \boldsymbol{p}, \sigma , \pm \rangle  = (-)^{j-\sigma} | - \boldsymbol{p}, -  \sigma, \pm \rangle$ and the S-matrix for the scattering of these states would satisfy \eqref{T-INVA-S-MATRIX} without having to change particle type. Usually, of course, \eqref{redefinition} can not be done because we can not prepare superpositions of states with different $U(1)$ charges. For time reversal doublets, though, any superposition of the particle states is allowed; indeed, we have assumed this from the beginning so that \eqref{generalact} can be transformed in a set of doublets plus a diagonal action. Regardless of the basis of one particle state we choose, the $S$ matrix will never satisfy \eqref{T-INVA-S-MATRIX}. The point is that even if the Lagrangian of a time reversal doublet is identical to that of a $CT$ invariant theory, the two possibilities could be distinguished by the observation that in a time reversal doublet, any superpositions of the states belonging to the doublet are possible. Yet, no basis redefinition can satisfy the S-matrix requirement \eqref{T-INVA-S-MATRIX}. We can contrast this with the standard discussion in \cite{Lee_Wick}. In that seminal paper, the main argument is that there is an inevitable ambiguity in the definition of space-time reflection operators. For instance, if the Lagrangian is invariant by two antilinear operators that transform fields at $x$ into a linear combination of fields at the point $Tx$, both can be called the time-reversal operator. Here, we call attention to the fact that this is not the case in the absence of selection rules (not considered in \cite{Lee_Wick}). The definition of the time reversal operator, the choice between $CT$ and a $T$ so that $T^2 \phi(x) T^{-2} \neq \phi(x)$, would be required by the existence of all possible superposition of states and the impossibility of choosing a basis so that \eqref{T-INVA-S-MATRIX} holds. 

Although \eqref{poss1} could represent a possible model for time reversal doublets, it contains a derivative coupling that deserves attention and, again, the model appears unsuitable due to renormalizability issues. Therefore, since we have to introduce two currents to break the symmetry under $S$, developing a renormalizable model of one scalar and one massive vector field does not seem possible. On the other hand, there is a simple extension of what we have done, providing a renormalizable model. Suppose we have two complex fields with the following space-time reflection properties 
\begin{align}
T \phi_i(x) T^{-1} = e^{i\phi/2} \phi^\dagger_i (-Px), \qquad P \phi_i(x) P^{-1} = \eta \phi_i(Px),
\end{align}
for $i=1,2$. We shall assume $U(1)$ invariance, and that the fields have the same charge, that is 
\begin{align}
\left[ Q \, , \, \phi_i \right] = q \, \phi_i, \qquad i=1,2.
\end{align}
Among the possible $U(1)$ invariant terms we have $\text{Re} \left[ \phi^*_1 \phi_2 \right] i \text{Im} \left[ \phi^*_1 \phi_2 \right]$, which is $T$ and $P$ invariant but not $S$ invariant, so that the model is not trivial. The full renormalizable Lagrangian would be 
\begin{align}
\mathcal{L} = \mathcal{L}_0  +  \lambda_1 (\phi_1 \phi^*_1)^2 + \lambda_2 (\phi_2 \phi^*_2)^2  + i\, \lambda_3 \,  (\phi_1 \phi^*_2 - \phi_2 \phi^*_1),
\end{align}
which is $U(1)$ invariant and contains only scalar fields, in which case we do not expect problems with renormalizability. One can also check that there are no $P$, $T$ invariant couplings of these fields and the photon. Naturally, one could go on and provide either more complicated interactions involving scalars or interacting models of spin half particles (or even both). 

\subsection{On the connection with dark matter}

Since standard matter, baryons and leptons, and their interactions break $P$ and $T$, previous considerations on degeneracy under space-time reflections assumed that, if they exist, they should be approximations \cite{Lee_Wick}. However, there are other possibilities. There is a scenario in which this degeneracy could be exact. We assume the existence of a set of particle states that undergo mixing under the action of $T$ and possibly $P$, and any scattering process involving these particles is invariant under $P$ and $T$. As the interactions of ordinary matter are not invariant under $P$, and $T$, there can be no coupling between the fields of these particles and the fields associated with standard matter\footnote{This is also true for possible mediators like $A_\mu$ in \eqref{poss1}. }. If such a coupling did exist, it would be possible to construct an $S$ matrix element involving doublets and standard matter that does not have a defined transformation under space-time reflections\footnote{All we need are connected diagrams where some vertices are invariant under space-time reflections and others that are not.}. Therefore, although this sector has self-interactions, it would not interact with the standard model, rendering it invisible under standard interactions. Even though this representation of space-time reflections does not interact with the standard model, all fields are coupled to the gravitational field. Naturally, one could detect the existence of this sector by its gravitational influence on the standard model sector. Therefore, such representations of space-time reflections provide a candidate for dark matter. Of course, in light of the exposition of the preceding sections, this idea can only maintain consistency if gravitational interactions do not introduce violations of space-time reflections on time-reversal doublets. 

Despite the generality, this line of investigation already provides immediate consequences. First, as mentioned, there are no interactions between both sectors apart from the gravitational one. Conversely, experimental verification of non-gravitational interactions between dark matter and standard model particles would immediately exclude this approach. Also, there must be some non-gravitational self-interaction between these dark matter field candidates since a free theory of doublets is always invariant by $S$. In other words, the analysis points to the modeling of self-interacting dark matter, which has been considered a possible solution for small-scale problems of cold dark matter \cite{SIDM}. Lastly, since our criteria to distinguish Wigner doublets from ordinary field theory relies on the absence of selection rules (which are not directly probed, in general), in order to identify dark matter with Wigner doublets, one would have to observe some indirect consequence of superpositions that are otherwise forbidden by selection rules.

\section{Concluding remarks}

From the theoretical point of view, we have studied a field whose states are endowed with degeneracy beyond the spin, severely constraining the construction via requirements coming from the very existence of non-standard Wigner classes and the theoretical viability of the models as dark matter candidates. We have arrived in a class of self-interacting, but otherwise invisible to standard model interaction models. Of course, additional quantum numbers labeling physical states have never been found in accelerators. However, in a context where data suggest that most of the universe's matter content is unknown, it seems reasonable to investigate if irreducible representations encompassing reflections may also serve as candidates for, at least, part of the dark matter sector.    

As previously mentioned, the present understanding of $P$ and $T$ symmetry in some sectors of the standard model suggests that these are accidental symmetries. Since the approach proposed in this paper reestablishes $P$ and $T$ as fundamental symmetries, it requires a division of the matter content in nature into two sectors, so to speak, with only one preserving the entire Lorentz group. Questions regarding why this split should occur or whether there is a mechanism that produces this division are beyond the scope of this paper.

\subsection*{Acknowledgements}
The authors mostly benefited from conversations with Dr. Gustavo P. de Brito. JMHS thanks to the National Council for Scientific and Technological Development -- CNPq (grant No. 307641/2022-8).

\end{document}